\documentstyle[twocolumn,prb,aps,epsf]{revtex}
\def\inseps#1#2{\def\epsfsize##1##2{#2##1} \centerline{\epsfbox{#1}}}
\begin{document}
\draft
\input{psfig}
%\twocolumn[\hsize\textwidth\columnwidth\hsize\csname @twocolumnfalse\endcsname] 
\title{Spinodal decomposition of  binary mixtures in  uniform shear flow}
\author{F. Corberi}
\address{Dipartimento di Scienze Fisiche, Universit\`a di Napoli and
Istituto Nazionale di Fisica della Materia, Unit\`a di Napoli,
Mostra d'Oltremare, Pad.19, 80125 Napoli, Italy}

\author{G. Gonnella and A. Lamura}
\address{Dipartimento di Fisica, Universit\`a di Bari,
Istituto Nazionale di Fisica della Materia, Unit\`a di Bari,
and
Istituto Nazionale di Fisica Nucleare, Sezione di Bari, via Amendola
173, 70126 Bari, Italy.}
\date{\today}
\maketitle
\begin{abstract}
Results are presented for the phase separation process
of a binary mixture subject to an uniform shear flow quenched 
from a disordered to a homogeneous ordered
phase. The kinetics of the process is described in the context of   
the time-dependent Ginzburg-Landau equation with an external
velocity term. The  one-loop approximation is used to study the evolution
of the model. We show that the structure factor obeys a generalized
dynamical scaling. The domains grow with different typical lengthscales
$R_x$ and  $R_y$ respectively 
in the flow and in the shear directions. In the 
scaling regime $R_y \sim t^{\alpha_y}$ and $R_x \sim  t^{\alpha_x}$,
 with $\alpha _x=5/4$ and
$\alpha _y =1/4$. 
The excess viscosity $\Delta \eta$ after reaching a maximum 
relaxes to zero as $\gamma ^{-2}t^{-3/2}$, $\gamma$ being the 
shear rate. $\Delta \eta$ and other
observables exhibit log-time periodic oscillations which can be
interpreted as due to a growth mechanism where 
stretching  and break-up of domains cyclically occur.
\end{abstract}


The kinetics of the growth of ordered phases as a disordered system is quenched
into a multiphase coexistence region has been extensively studied
in the last years [\onlinecite{Bin}]. The main features
 of the process of phase separation are well understood.
Typically, domains of the ordered phases grow with the law 
$R(t) \sim t^{\alpha}$,
where $R(t)$ is a measure of the average  size of domains.
The pair correlation function $C(r,t)$ verifies 
asymptotically a dynamical scaling
law according to which it can be written as $C(r,t) \simeq f(r/R)$, 
where $f(x)$ is a scaling function. 
In particular, in binary liquids,  
the existence of various regimes characterized
by different growth exponents $\alpha$ is well established [\onlinecite{B94}].
In this letter we study the process of phase separation in a binary
mixture subject to an uniform shear flow.  
When a shear flow is applied to the system, the growing domains are affected 
by the flow and the time evolution is substantially different from that of
ordinary spinodal decomposition [\onlinecite{On97}]. 
The scaling behavior of such a system is not clear. 
Here we show the 
existence of a scaling theory with 
different growth exponents for the flow  and  the other directions.
For long times, in the scaling regime,
the observables are modulated by {\it log-time periodic} oscillations
which can be related to a mechanism of storing and dissipation
of elastic energy. 
The  behavior of the excess viscosity and other rheological indicators 
reflects this mechanism and is also calculated.

The problem is addressed in the context of the time-dependent
Ginzburg-Landau equation for a diffusive  field coupled with  an external
velocity field [\onlinecite{On97}].  
The binary mixture is described by the equilibrium
free-energy
\begin{equation}
{\cal F}\{\varphi\} = \int d^d x 
\{\frac{a}{2} \varphi^2 + \frac{b}{4} \varphi^4 
+ \frac{\kappa}{2} \mid \nabla \varphi \mid^2 \}
\label{eqn1}
\end{equation}
where $\varphi$ is the order parameter describing the concentration
difference between the two components. The parameters $b,\kappa $ are 
strictly positive in order to ensure
stability; $a<0$ in the ordered phase. 
The Langevin equation for the evolution of the system is
\begin{equation}
\frac {\partial \varphi} {\partial t} + \vec \nabla (\varphi \vec v) =
\Gamma \nabla^2  \frac {\delta {\cal F}}{\delta \varphi} + \eta
\label{eqn2}
\end{equation}
where $\eta$ is a gaussian stochastic field representing the effects 
of the temperature in the fluid. The fluctuation-dissipation
theorem requires that
\begin{equation}
<\eta(\vec r, t) \eta(\vec r', t')> = -2 T \Gamma \nabla^2 \delta(\vec r - 
\vec r') \delta(t-t')
\label{eqn3}
\end{equation}
where $\Gamma$ is a mobility coefficient, $T$ is the 
temperature of the fluid, and the symbol $<...>$ denotes the 
ensemble average.
We  consider the simplest shear flow with velocity profile given by 
\begin{equation}
\vec v = \gamma y \vec e_x
\label{eqn4}
\end{equation}
where $\gamma$ is the spatially homogeneous shear rate [\onlinecite{On97}]
and $\vec e_x$ is a unit vector in the flow direction.

In the  process of phase separation 
the  initial configuration of $\varphi$ is  
a high temperature disordered state and
the evolution of the system is studied in  model  (\ref{eqn2}) with $a<0$.
It is well-know that in this case, also without the velocity term, 
the model (\ref{eqn2}) cannot be solved exactly [\onlinecite{B94}].
In this letter we deal with 
the non-linear term of eq. (\ref{eqn2}) in the one-loop
approximation [\onlinecite{MZ,PD}]. 
In this approximation the term $\varphi^3$ appearing
in the derivative $\delta {\cal F}/\delta \varphi$ is linearized as
$<\varphi^2> \varphi$. It is also called large-n limit. Indeed, in the case
of a vector field  $\vec \varphi$ with $n$-components the term
$(\vec \varphi^2) \vec \varphi$ reduces to $<\varphi^2> \varphi$
in the $n \rightarrow \infty $ limit [\onlinecite{Ma}]. 
The validity and the limitations 
of this approximation, due to the acquired vectorial 
character of  the order parameter,
are discussed  in literature [\onlinecite{CCZ}].

Before presenting our results it is useful to summarize the known
behaviour of a phase separating mixture under shear flow. The shear induces 
strong deformations of the  bicontinuous pattern appearing
after the quench [\onlinecite{On97,OND,Ro}]. 
When the shear is strong enough stringlike
domains have been observed to extend macroscopically
in the direction of the flow [\onlinecite{Has}]. 
In experiments a value 
$\Delta \alpha  = \alpha_x  - \alpha_y$ in the range $0.8\div 1$ 
for the difference 
of the growth exponents in the flow and in the shear directions is measured
[\onlinecite{LLG,Bey}].
Two dimensional molecular dynamic simulations 
find a slightly smaller value  
[\onlinecite{PT}].
We are not aware of any existing theory for the value of $\alpha_x, 
\alpha_y $.
The shear also induces a peculiar rheological behavior.
The break-up of the stretched domains liberates an energy which gives rise
to an increase $\Delta \eta$ of the viscosity  
[\onlinecite{Onu,KSH}]. 
Experiments and simulations show that the
excess viscosity $\Delta \eta$ reaches a maximum at $t=t_m$ 
and then relaxes to smaller values.
The maximum of the excess viscosity is expected to occur at a fixed 
$\gamma t$  and to scale as  
$\Delta \eta (t_m) \sim \gamma^{-\nu}$ [\onlinecite{OND,LLG}].
Simple scaling arguments 
predict $\nu=2/3$ [\onlinecite{OND}], but different values 
have been reported [\onlinecite{LLG}]. 

We  study the time evolution of the structure factor
\begin{equation}
C(\vec k,t) = <\varphi(\vec k, t) \varphi (-\vec k,t)>
\label{eqn5}
\end{equation}
where $\varphi(\vec k,t)$ is the Fourier transform of the
field $\varphi(\vec x, t)$ solution of  eq. (\ref{eqn2}).
The excess viscosity is defined in terms of $C(\vec k,t)$
by 
\begin{equation}
\Delta \eta (t)= -\gamma^{-1} \int _{|\vec k |<q}
\frac {d\vec k}{(2\pi)^D} k_x k_y C(\vec k,t) 
\label{eqn5b}
\end{equation}
where $q$ is a phenomenological cutoff.
In the one-loop approximation the dynamical equation for 
$C(\vec k,t)$ is:
\begin{equation}
\frac {\partial C(\vec k,t)} {\partial t} - 
\gamma k_x \frac {\partial C(\vec k,t)} {\partial k_y} =    
- k^2 [k^2 + S(t) -1]C(\vec k,t) + k^2 T
\label{eqn6}
\end{equation}
where
\begin{equation}
S(t)  =  \int _{|\vec k|<q}  \frac {d\vec k}{(2\pi)^D}  C(\vec k,t)
\label{eqn7}
\end{equation}
The parameters $\Gamma, a, b, \kappa$ have been eliminated by a redefinition
of the  time, space and field scales. We solve Eq.(\ref{eqn6}) numerically 
in two dimensions.
A first-order Euler scheme is implemented with an 
adaptive mesh, due to the peaked character of the solution.
The initial condition chosen for the function $C(\vec k,0)$ is a constant 
value, which corresponds to the  disordered state with $T=\infty$.
The  typical evolution of $C(\vec k,t)$ is shown  in Fig.1 
for the particular case  $T=0$ and  $\gamma=0.001$.
At the beginning the function $C(\vec k,t)$ evolves forming
a circular volcano structure, as usually in the case without shear. 
This is the early-time regime when well-defined
domains are forming.
Then shear-induced anisotropy
effects become evident in the elliptical shape of  $C(\vec k,t)$ and in the 
profile of the edge of the volcano, 
as it can be seen in Fig.1 at $\gamma t=0.05$. 
Similar elliptical patterns of $C(\vec k,t)$ are  usually observed in 
experiments. 
The dips in the edge of the volcano develop with time until $C(\vec k,t)$
results  to be separated in two distinct foils, as $\gamma t\simeq 1$.
This explains the disappearing of the peak corresponding to the major axis
of the ellipse observed in experiments [\onlinecite{Bey}]. 
During this evolution the
support of $C(\vec k,t)$ 
shrinks towards  the origin
with different scales for the shear and  the flow directions. 
At later times  in 
each foil  of $C(\vec k,t)$  two peaks can be distinguished 
and the relative heights of
these peaks change in time. In Fig.1 at $\gamma t=6$ 
the peak characterized by $|k_y| \gg |k_x|$ dominates, 
while the other peak with $|k_y|\simeq |k_x|$ 
prevails successively. The oscillations
between the two peaks have been observed to continue in time and characterize
the steady state. 

\begin{figure}
\vskip -3mm
\inseps{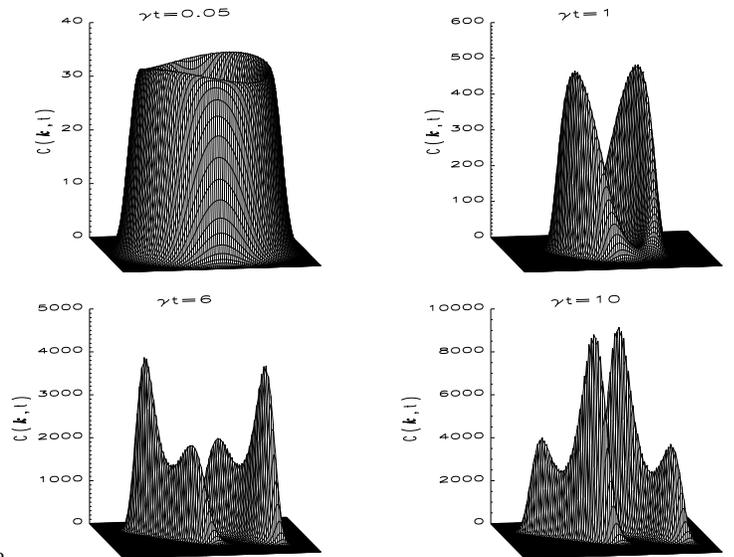}{0.60}
\vskip +5mm
\caption{The structure factor at consecutive times for $\gamma=0.001$.
The  $k_x$ coordinate is on the horizontal axis and assumes positive 
values on the right of the pictures,  while the $k_y$ is positive
towards the upper part of the coordinate plane. 
The support of the function $C(\vec k,t)$ 
shrinks towards the origin.
For a better view of $C(\vec k,t)$, in the last two pictures,
 we have enlarged differently the scales on the $k_x$ and $k_y $ axes.
The actual angle between the direction of the foils of $C(\vec k,t)$
and the $k_y$ axes is $\theta = 21^o$ and $\theta = 13^o$ in the last two
pictures.} 
\end{figure}

A quantitative measure of the size of domains is given by $R_x(t)= 
1/\sqrt {<k_x^2>}$ where 
$ <k_x^2> = \int d\vec k k_x^2 C(\vec k,t) / \int d\vec k  C(\vec k,t)$,
and the same for the other directions.
The evolution of $R_x, R_y$  is plotted in Fig.2.
The growth exponents in the shear and in the flow direction
are  $\alpha_y \simeq1/4$ and $\alpha_x \simeq 5/4$. 
The value $\alpha_y =1/4$ is the same as
in models with vectorial conserved order parameter without shear; 
this corresponds to
the Lifshitz-Slyozov exponent $\alpha =1/3$ for scalar fields. 
A growth exponent $\alpha _y $
unaffected by the presence of shear is also measured  in experiments 
[\onlinecite{LLG}].  
We see in Fig.2 that the amplitudes of  $R_x, R_y$         
oscillate periodically in  logarithmic time. This behavior
 can be related to the oscillations of  the peaks of  
$C(\vec k,t)$ previously observed  and will be discussed later 
in relation with the behavior of the excess viscosity.

\begin{figure}
\vskip -3mm
\inseps{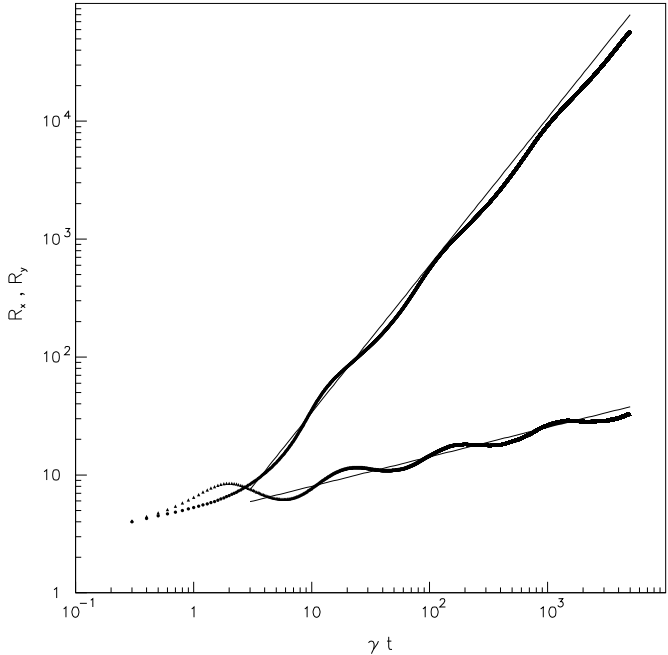}{0.5}
\vskip -10mm
\caption{The average size of domains in the 
$x$ and $y$ directions as a function of the strain $\gamma t$.
The two straight lines have slope 5/4 and 1/4.}
\end{figure}

In order to study analytically the behavior of the model 
for arbitrary space dimensionality $d$ we
resort to a scaling ansatz [\onlinecite{note}]. For the structure
factor we then assume
\begin{equation}
C(\vec k,t)=\prod _{i=1} ^d R_i (t)F\left [\vec X,
\tau(\gamma t)\right]
\label{scaling}
\end{equation}
for long times, where $\vec X$ is a vector of
components $X_i=k_i R_i(t)$, $F$ is a scaling function and the subscript $i$
labels the space directions with $i=1$ along
the flow.
We also allow an explicit time dependence
of the structure factor through $\tau(\gamma t)$; notice that since
$C(\vec k,t)$ scales as the domains volume below the critical temperature,
$\tau $ must not introduce any further algebraic
time dependence in $C(\vec k,t)$. We then
argue that $F$ is a periodic function of $\tau$,
as suggested by the oscillations observed numerically
in the physical observables.
Inserting this form of $C(\vec k,t)$
into eq.~(\ref{eqn6}) we obtain:
\begin{eqnarray}
\gamma X_1F_2
    &=& 
           R_1 R_2^{-1} \left \{ 
           \dot \tau \partial F /\partial \tau +\sum _{i=1} ^d
           \Bigg [ R_i^{-1}\dot R_i (F+X_iF_i)
           + \right .  \nonumber \\
    & &    \left .  R_i^{-2}X_i^2 \left ( \sum _{k=1} ^d
           R_k^{-2}X_k^2 -1+S \right ) F \Bigg ] \right \}
\label{ciccio}
\end{eqnarray}
where $F_i=\partial F/\partial X_i$ and a dot means
a time derivative.
Since the l.h.s. of Eq.(\ref{ciccio}) scales as $t^0$ 
one has the solution $R_i(t)\sim
\gamma ^{\delta _i} t^{\alpha _i}$, $\tau (\gamma t)\sim \log \gamma t$,
$S(t)=1-t^{-\beta}$, with $\delta _1=1$, $\delta _i =0$
$(i=2,d)$, $\alpha _1=5/4$, $\alpha _i=1/4$ $(i=2,d)$
and $\beta =1/2$. In this way we recover the growth exponents 
previously found. Actually the exponents found numerically
are slightly smaller then the predicted powers due to logarithmic
corrections [\onlinecite{note}]. 

We now turn to the analysis of the rheological behavior of the mixture
and in particular of the excess viscosity. 
The previous theoretical
arguments can be used to establish the scaling properties of $\Delta \eta$.
Inserting the form~(\ref{scaling}) into Eq.~(\ref{eqn5b})
we obtain $\Delta \eta (t)\sim \gamma ^{-1}R_1(t)^{-1}R_2(t)^{-1}g(\tau)
\sim \gamma ^{-2}t^{-3/2}g(\tau)$, where
$g(\tau)=\int X_1 X_2 F \left [ \vec X,\tau (t)\right ]d\vec X$
is a periodic function of $\tau (\gamma t)$. Therefore,  in the scaling
regime, for each value of $\gamma t$,   
the functions $\Delta \eta$ corresponding to
different values of $\gamma$ collapse each on the others if rescaled as
$\Delta \eta \rightarrow \gamma ^{1/2} \Delta \eta$. 
A similar 
analysis can be done for the normal stress which 
is  defined as  $ \Delta N_1 =  \int \frac {d\vec k}{(2\pi)^D} 
[k_y^2- k_x^2] C(\vec k,t)$ and  scales as $t^{-1/2}$.

The  behavior of  $\Delta \eta$ at all times, calculated by the 
numerical expression of  $C(\vec k, t)$, is shown  in Fig.3  
for the case $\gamma=0.001$.
$\Delta \eta$ reaches a maximum at $\gamma t \simeq 3.5$, 
then it decreases 
with the power law $t^{-3/2}$ modulated by a  periodic oscillation in 
logarithmic time. 
A comparison with Fig. 2 shows that 
the asymptotic scaling regime starts when 
the excess viscosity
reaches its maximum at $t=t_m$, as found also in experiments 
[\onlinecite{LLG}]. 
 The occurrence of the predicted scaling of $\Delta \eta$ with $\gamma$
is verified numerically with great accuracy for long times. 
However, since $t_m$ is at the onset of scaling,
an effective exponent somewhat larger then $1/2$ 
($\nu \simeq 0.6$) is measured 
for $\Delta \eta (t_m)$, due to preasymptotic corrections. 

The periodic oscillations of $\Delta \eta$ 
are due to the competition between the different peaks of  $C(\vec k,t)$.
A local maximum of $\Delta \eta$ occurs for a situation 
similar to that of  Fig.1 at $\gamma t =6$, when the peak 
with $|k_y|\gg |k_x|$ dominates and the difference between
the height of the two peaks is maximal. The minima of $\Delta \eta $
correspond to the opposite situation, as in Fig.1 at $\gamma t=10$.
The oscillations can be explained in this way: 
The elongation of the domains in the flow
direction produces an increase of $\Delta \eta$. 
Stretched domains are characterized by 
$R_y \ll R_x$ and, therefore, are represented by the peak of $C(\vec k,t)$
with $|k_y|\gg |k_x|$, which dominates in this time domain. 
As time passes, however, domains are deformed
to such an extent that they start to burst dissipating the stored energy.
As a consequence $\Delta \eta$ decreases and more isotropic domains
are formed. These are characterized by similar values of $R_x$ and $R_y$
and correspond to the other peak of $C(\vec k,t)$.
This peak starts growing faster than the other until it prevails.
Later on a minimum of $\Delta \eta$ is observed. Then elongation
occurs again and this mechanism reproduces periodically in time
with a characteristic frequency. To our knowledge the existence of
this periodic behavior has never been discussed before [\onlinecite{note1}].

\begin{figure}
\vskip -3mm
\inseps{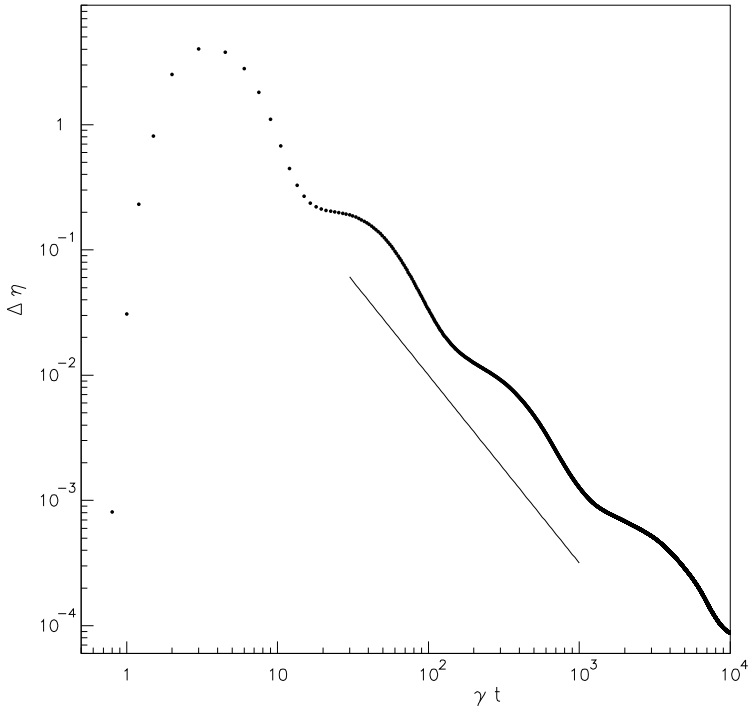}{0.55}
\vskip -13mm
\caption{The excess viscosity 
as a function of the strain
$\gamma t$. The slope of the straight line is -3/2.}
\end{figure}

To conclude, we have studied the phase separation of a binary mixture
in an uniform shear flow. Dynamical scaling holds for this system.
Domains grow along the flow as $R_x(t) \sim t^{5/4}$
while in the other directions the exponent of the diffusive growth
is the same as without shear. The difference $\Delta \alpha$
between the growth exponents is 1, a result which is consistent with
real experiments.
The excess viscosity after the  maximum
relaxes to zero as  $\gamma ^{-2}t^{-3/2}$. The amplitudes of physical
quantities are decorated by oscillation periodic in logarithmic time. 
It would be interesting to study these phenomena in direct simulation
of the Langevin equation and also
to see the effects of hydrodynamics on this system. 
 
~\\
We thank Julia Yeomans for helpful discussions. 
F.C. is grateful to M.Cirillo for hospitality in the University of Rome.
F.C. acknowledges support by the TMR network contract ERBFMRXCT980183.


\begin{references}

%4

\bibitem{Bin}
K. Binder in 
``Phase Transitions in Materials'', Materials Science and Technology Vol. 5,
eds. R.W. Cahn, P. Haasen, and E.J. Kramer (VCH Weinheim 1990);
H. Furukawa,  Adv. in Phys.
{\bf 34}, 703 (1985); J. D. Gunton {\it et al. }, in ``Phase Transitions
and Critical Phenomena'', Vol. 8, eds. C. Domb and J.L. Lebowitz 
(Academic 1983).

\bibitem{B94}
A.J. Bray, Adv. in Phys. {\bf 43} 357 (1994). 

%5
\bibitem{On97}
For a review, see A. Onuki, J. Phys.: Condens. Matter {\bf 9} 6119 (1997).

%6a
\bibitem{MZ}
G.F. Mazenko and M. Zannetti, Phys. Rev. Lett. {\bf 53}, 2106 (1984);
Phys. Rev. B {\bf 32}, 4565 (1985).

%6b

\bibitem{PD}

G. P\"{a}tzold and K. Dawson, Phys. Rev. E {\bf 54}, 1669 (1996). 

%7
\bibitem{Ma}
See, e.g., S.k. Ma in ``Phase Transitions
and Critical Phenomena'', Vol. 6, eds. C. Domb and M.S. Green 
(Academic 1976)


%8 
\bibitem{CCZ}
C. Castellano, F. Corberi, and M. Zannetti, Phys. Rev. E {\bf 56}, 4973 (1997).

%8a
\bibitem{OND}
T. Ohta, H. Nozaki, and M. Doi, Phys. Lett. A {\bf 145} 304 (1990);
J. Chem. Phys. {\bf 93} 2664 (1990).

%8b
\bibitem {Ro}
D.H. Rothman, Europhys. Lett. {\bf 14} 337 (1991).


%9
\bibitem{Has}
T. Hashimoto, K. Matsuzaka, E. Moses, and A. Onuki, Phys. Rev. Lett. {\bf 74} 
126 (1994). 

%10
\bibitem{LLG}
J. L\"{a}uger, C. Laubner, and W. Gronski, Phys. Rev. Lett. {\bf 75} 
3576 (1995).

%10a
\bibitem{Bey}
C.K. Chan, F. Perrot, and D. Beysens, Phys. Rev. A {\bf 43} 
1826 (1991).

%11
\bibitem{PT}
P. Padilla and S. Toxvaerd, J. Chem. Phys. {\bf 106} 2342 (1997).

%13
\bibitem{Onu}
A. Onuki,  Phys. Rev. A {\bf 35} 5149 (1987).

%12
\bibitem{KSH}
A.H. Krall, J.V. Sengers, and K. Hamano, Phys. Rev. Lett. {\bf 69} 
1963 (1992).

%14a
\bibitem{note}
It is well known 
[A.Coniglio, P.Ruggiero and M. Zannetti, Phys. Rev. {\bf E} 50,
1046, (1994)] that in the present approximation
simple scaling is not obeyed for $\gamma =0$ and $C(\vec k,t)$, 
instead of scaling
with the domains volume as in Eq.(\ref{scaling}), shows a
continuum spectrum of $\vec k$-dependent exponents
(multiscaling). However standard scaling is the leading order
approximation in the region surrounding the peak of the structure factor.
With a simple scaling ansatz, therefore, one obtains the correct
value of the growth exponents (apart from logarithmic corrections)
because the peak contribution dominates the momentum integrals
that define the physical observables. Since we do not have
presently an exact solution with $\gamma \neq 0$ multiscaling cannot be,
in principle, ruled out but the same considerations applies in the
peaks regions. Furthermore simple scaling is expected when the
present approximation is released 
[A.J.Bray and K.Humayun, Phys. Rev. Lett. {\bf 68}, 1559 (1992)].
                        
%15
\bibitem{note1}
A logarithmic-time periodic release of elastic energy
has been also observed in models for the propagation of fractures in 
materials subject to an external strain (M. Sahimi and S. Arbabi, Phys. Rev.
Lett. {\bf 77} 3689 (1996)). See also D. Sornette, Phys. Rep. {\bf 297}, 
239 (1998).

\end{references}
\end{document}